\documentclass[journal]{IEEEtran}

\usepackage{amssymb}
\usepackage{amsmath,amssymb,amsthm,color}
\usepackage{graphicx}
\usepackage{subfigure}
\usepackage{epsfig}
\usepackage{setspace}
\usepackage{cases}
\usepackage{tabularx, multirow, booktabs}
\usepackage{wasysym}
\usepackage{booktabs}
\usepackage[ruled,vlined]{algorithm2e}
\usepackage{textcomp}

\IEEEoverridecommandlockouts

\begin{document}

\bibliographystyle{IEEEtran}

\title{Global Adaptive Routing Algorithm Without Additional Congestion Propagation Network}
\author{Shaoli Liu, Yunji Chen, Tianshi Chen, Ling Li, Chao Lu\\Institute of Computing Technology, Chinese Academy of Sciences}

\maketitle

\IEEEpeerreviewmaketitle

\section{Introduction}\label{section:Introduction}

\IEEEPARstart{A}{daptive} routing algorithm has been employed in
multichip interconnection networks in order to improve network
performance. Does a algorithm use local or global network state?
This is the key question in adaptive routing. In many traffic
patterns, the ignorance of global network state, leading to routing
selection based only on local congestion information, tends to
violate global load balance. To attack the load balance issue in
adapting routing, some global adaptive routing algorithms introduce
a congestion propagation network to obtain global network status
information, such as Regional Congestion Awareness (RCA) \cite{RCA}
and Destination Based Adaptive Routing (DBAR) \cite{DBAR}.

However, the congestion propagation network leads to additional
power and area consumption which cannot be ignored. From another
view, if we just increase the bandwidth between neighbor nodes with
the wires used to build the congestion propagation network, the
network performance could be improved as well. In this paper, we
propose a global adaptive routing algorithm without employing the
additional congestion propagation network. Our algorithm obtains the
global network state in a novel way, and can offer significant
improvement than the base-line local adaptive routing algorithm
(xy-adaptive algorithm which selects routing based on local
congestion information in each hop) for both medium and high
injection rates.

In wormhole flow control, all the routing information (flit id,
source node id, destination node id, vc id and address) is contained
in head flit, and data is carried in body flits. As a result, there
are always many free bits in the head flit, especially when the
bandwidth is 128-bits which is normal in interconnection network
design. Then, we can use these free bits in the head flit to
propagate global congestion information but not increase the number
of flits.

\section{Related Work}\label{section:Related}

Oblivious routing, in which the packets are routed without regard
for the network congestion state, is simple to implement and analyze
\cite{}. It is straightforward to compute the ideal, worst and
average case behavior of the oblivious routing algorithm on any
traffic pattern \cite{}.

An adaptive routing algorithm selects among alternative paths to
deliver a packet, by using information of the network congestion
state, typically virtual channel occupancies \cite{}. It has already
been successfully used in many commercial multi-core processors
\cite{}.

Theoretically, a good adaptive routing algorithm should have better
performance than an oblivious routing algorithm, since the
interconnection networks often have burst injection rates \cite{}
and the network congestion state information which could only be
known at run time is not available to the oblivious algorithm.
However, practically, many adaptive routing algorithms have poorer
worst-case performance than oblivious algorithm \cite{}. This is
largely because of the local nature of these adaptive routing
algorithms, that they just use local network congestion state when
making a routing decision. As a result, this shortsighted manner
which balances local load often results in global imbalance.

Regional Congestion Awareness (RCA) is the first algorithm to solve
the shortsighted problem of adaptive routing algorithm by utilizing
the non-local congestion state \cite{}. To attack the global load
balance issue, the authors present a congestion propagation
mechanism by employing an additional congestion propagation network.
However, mechanism used in RCA introduces redundant congestion
information in congestion calculation, which significantly reduce
the quality of congestion awareness.

\begin{figure}[htbp!]
\begin{center}
\epsfig{file=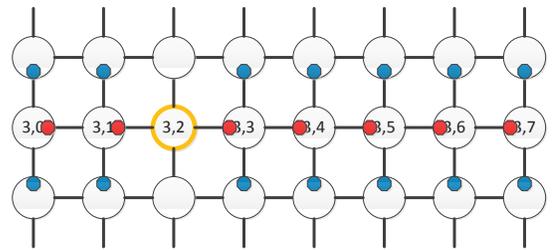,width = 0.4\textwidth}
\end{center}
\caption{Congestion information propagated by the additional
network.} \label{fig:insufficient}
\end{figure}

In order to eliminate excess congestion information,
Destination-Based Adaptive Routing (DBAR) employs a congestion
information propagation network, by which each router forwards the
number of available VCs to other routers in the same dimension.
While, because of the restriction of the wire-width of the
congestion information propagation network connecting neighbor
routers, insufficient congestion information is propagated. As shown
in Fig. \ref{fig:insufficient}, in horizontal dimension, only the
congestion states of the red ports (E(3,0) E(3,1) W(3,3) W(3,4)
W(3,5) W(3,6) and W(3,7)) are propagated to router (3,2), each port
one bit. While, we found in our experiments, the congestion states
of the blue ports in Fig. \ref{fig:insufficient} are also very
useful for the routing decision of node (3,2) in horizontal
dimension, which are not propagated to router (3,2) in DBAR
algorithm.

In our adaptive routing algorithm, we send the congestion
information without employ the congestion propagation network which
leads to additional power and area consumption that can not be
ignored. Furthermore, we propagate much more sufficient congestion
information than the DBAR algorithm, which leads to significant
improvement. And Our proposed algorithm provides deadlock avoidance
based on Duato¡¯s theory \cite{}.

\section{Algorithm}\label{section:Algorithm}

We will introduce our global adaptive routing algorithm in two
steps:
\begin{enumerate}
\item[$\bullet$]How to propagate global congestion information.
\item[$\bullet$]How to use global congestion information.
\end{enumerate}
We restrict our algorithm to mesh topology and minimal routing, but
the general ideas presented in this paper could be applied to other
topologies and non-minimally routing as well.

\begin{figure}[htbp!]
\begin{center}
\epsfig{file=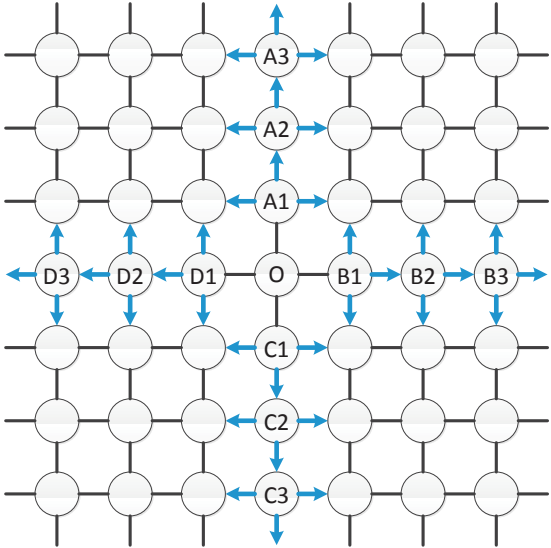,width = 0.4\textwidth}
\end{center}
\caption{The global congestion information (the blue arrows express
the direction of the congestion information) stored by node O.}
\label{fig:store}
\end{figure}

\begin{figure}[htbp!]
\begin{center}
\epsfig{file=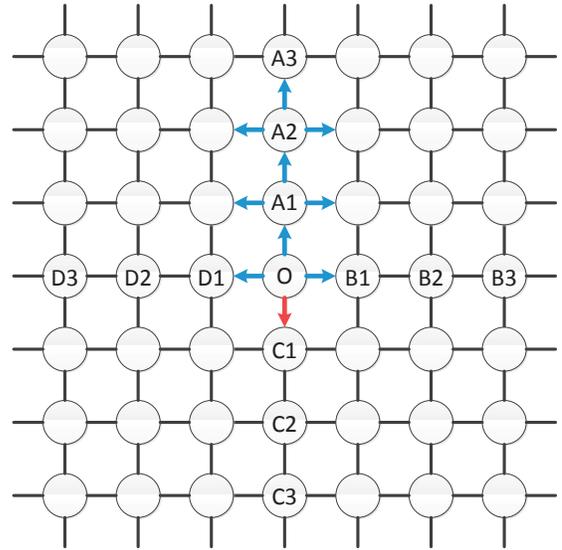,width = 0.4\textwidth}
\end{center}
\caption{The global congestion information (the blue arrows) carried
by the packets send form down port (the red arrow) of node O.}
\label{fig:carry}
\end{figure}

\subsection{Congestion Information}\label{subsection:Information}

As shown in Fig. \ref{fig:store}, the node O collects and stores the
congestion information of the nodes in the same col or row. Because
the congestion information of nodes far away form node O is useless,
we just look ahead as far as three hops. And take node A2 as an
example, since the congestion information of the down port of node
A2 is not useful for node O, node O only stores the congestion
information of the other three ports.

In our experiments, we use only 1 bits to express the congestion
information. We set the virtual channel number of each port as 8,
and if more than 4 virtual channels are occupied, the congestion
information is set to 1, else to 0. This is not just to save bits,
and we found in our experiments, 1 bits could get better performance
than 3 bits.

As show in Fig. \ref{fig:carry}, each time node O sends a packet
from a port (take down port as an example, the red arrow in Fig.
\ref{fig:carry}), we put the congestion information of three other
ports of node O and the congestion information of node A1 and A2
collected by node O (the blue arrows in Fig. \ref{fig:carry}) in the
head flits. We only use 9 free bits in the head flits, so the amount
of flits is not increased. And each time receiving a head flit, node
O updates the congestion information table with the congestion
information carried by it.

\subsection{Routing}\label{subsection:Routing}

\begin{figure}[htbp!]
\begin{center}
\subfigure[]{ \label{fig:subfig:dest}
\epsfig{file=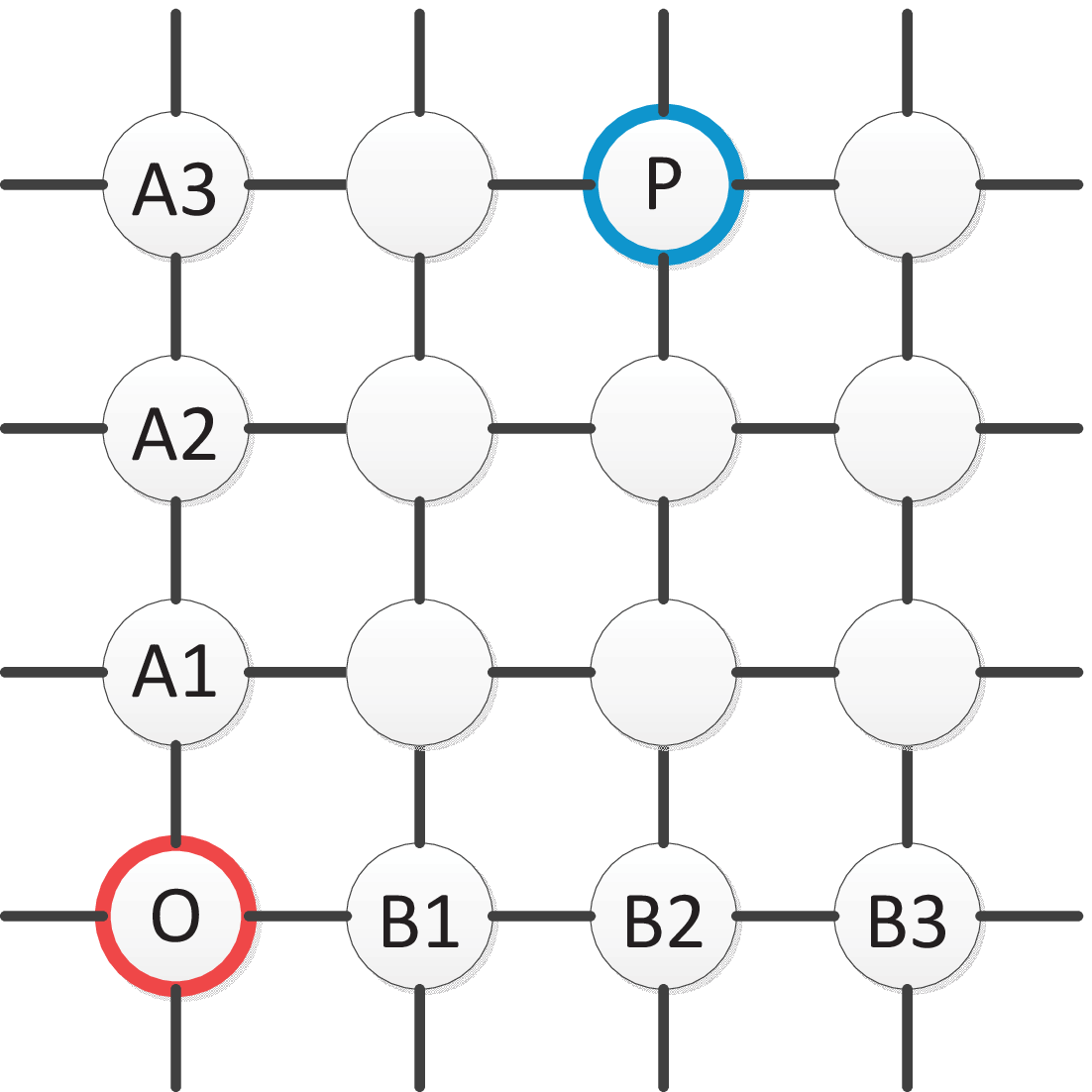,width = 0.2\textwidth}}\hspace{0.1in}
\subfigure[]{ \label{fig:subfig:step1}
\epsfig{file=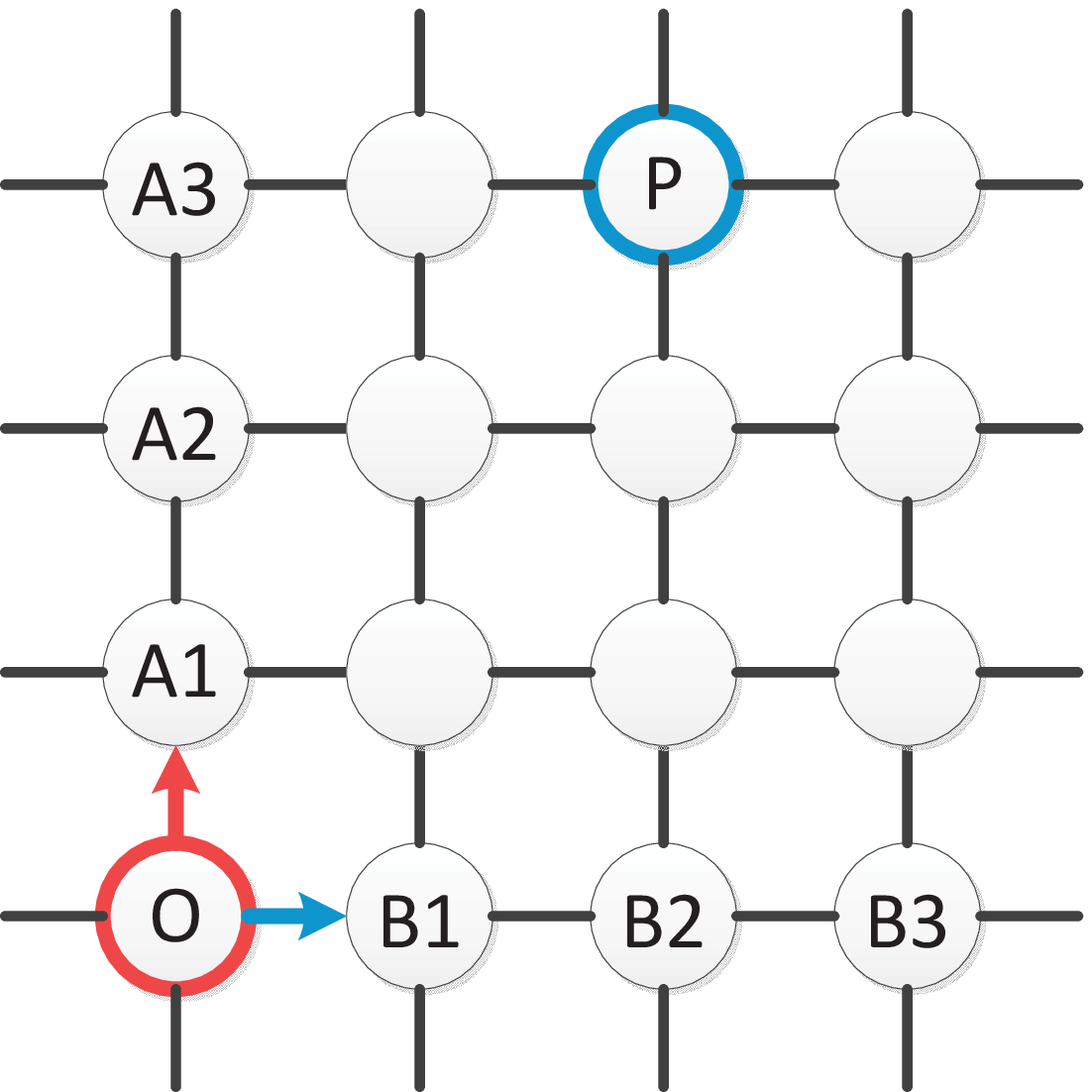,width = 0.2\textwidth}} \\
\subfigure[]{\label{fig:subfig:step2}
\epsfig{file=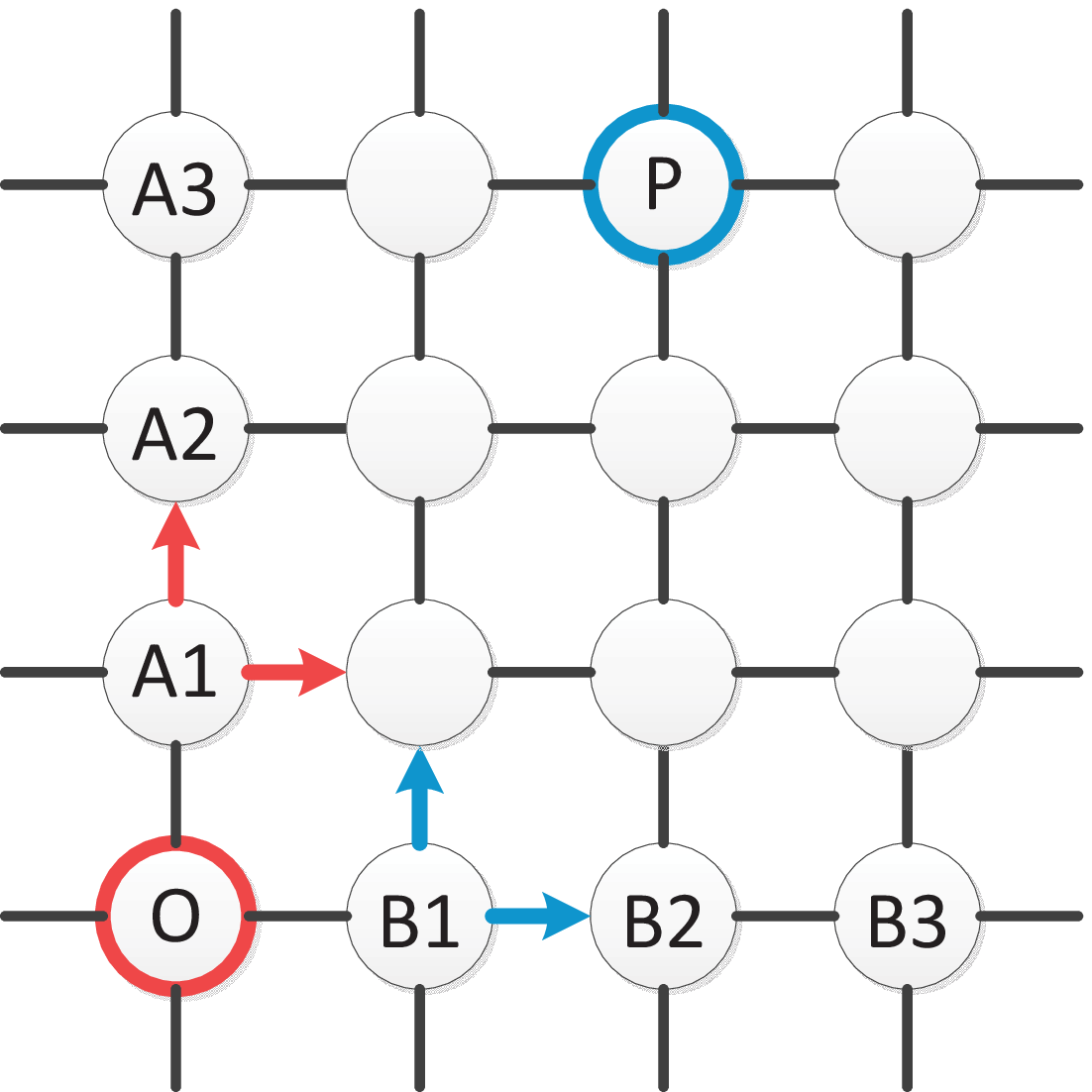,width = 0.2\textwidth}}\hspace{0.1in}
\subfigure[]{ \label{fig:subfig:step3}
\epsfig{file=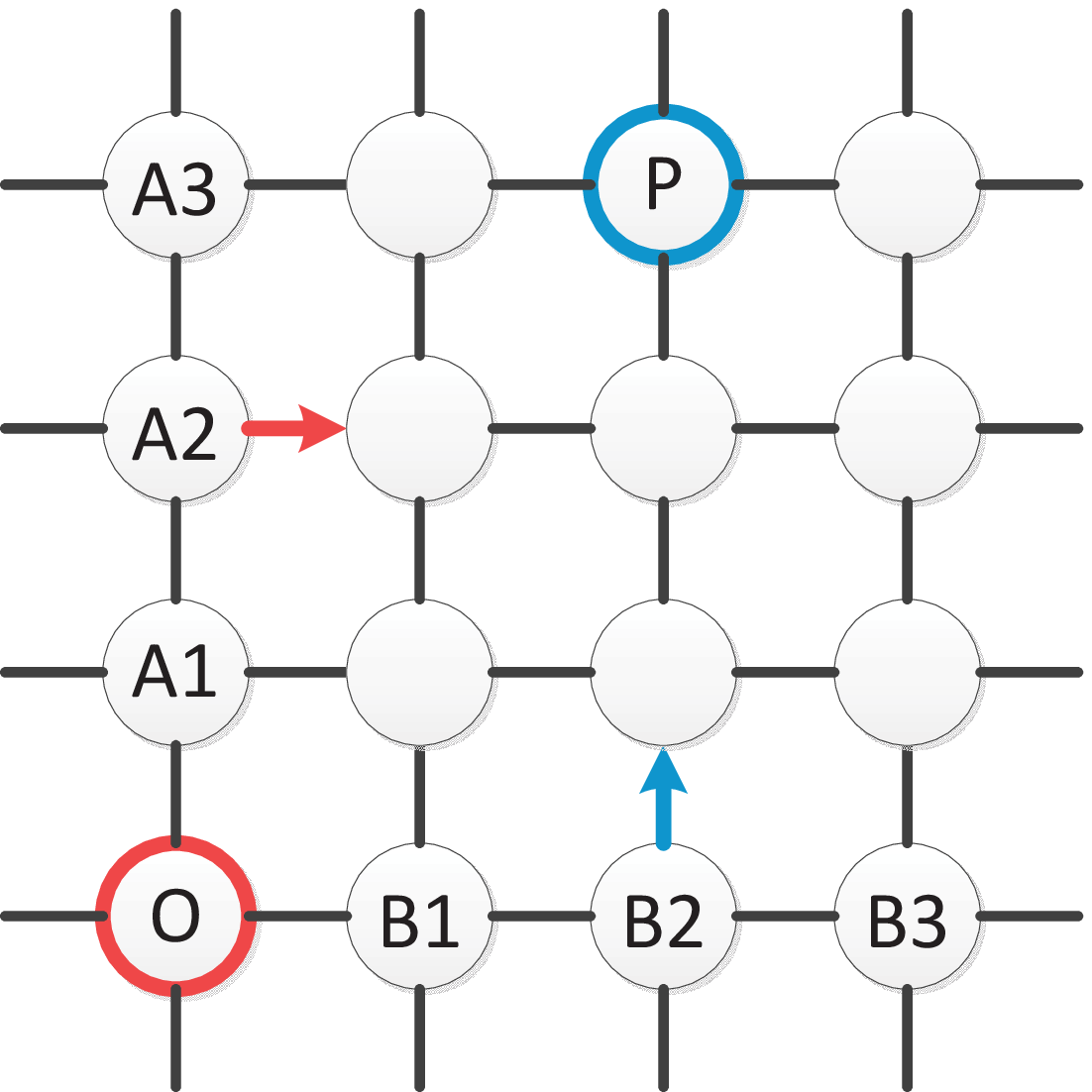,width = 0.2\textwidth}}
\end{center}
\caption{An example of our routing algorithm. (a) destination. (b)
step 1. (c) step 2. (d) step 3.} \label{fig:step}
\end{figure}

As shown in Fig. \ref{fig:subfig:dest}, a packet at node O need to
be sent to node P. First, as shown in Fig. \ref{fig:subfig:step1},
we compare the 1 bit congestion information of up port (red arrow)
and right port (blue arrow) of node O. If the congestion information
bits of the two ports are not equal, then take the direction with
smaller congestion information bit as the out direction and the
routing algorithm is end. Otherwise, look ahead one hop in each
direction as shown in Fig. \ref{fig:subfig:step2}. We add congestion
information of up port and right port of node A1 (red arrows) and B1
(blue arrows) respectively, and compare the two sum in the same way
as the step 1. If the routing algorithm is not end in step 2 either,
then we look ahead one more hop until reach the border (because we
use minimal routing, border means the farthest hop could be
transmitted in a direction) in any direction. As shown in Fig.
\ref{fig:subfig:step3}, B2 is the border of the right direction.
Because the right port of node B2 can not be used by this packet, we
only compare the congestion information of up port of B2 (blue
arrow) and right port of A2 (red arrow). If the congestion
information are always the same until we reach a border, then we
will take a random direction as the output.

\section{Experiment Results}\label{section:Results}

We use the same simulator (booksim) and experimental environment
(8VCs each port with 5flit buffers each VC, 8¡Á8 mesh topologies,
packet length is uniformly distributed between 1 and 6 flits,
128bits wire width) as the paper DBAR \cite{DBAR} used. But now we
only have the results of synthetic traffic patterns, because we do
not have application traces.



\begin{figure*}[htbp!]
\begin{center}
\epsfig{file=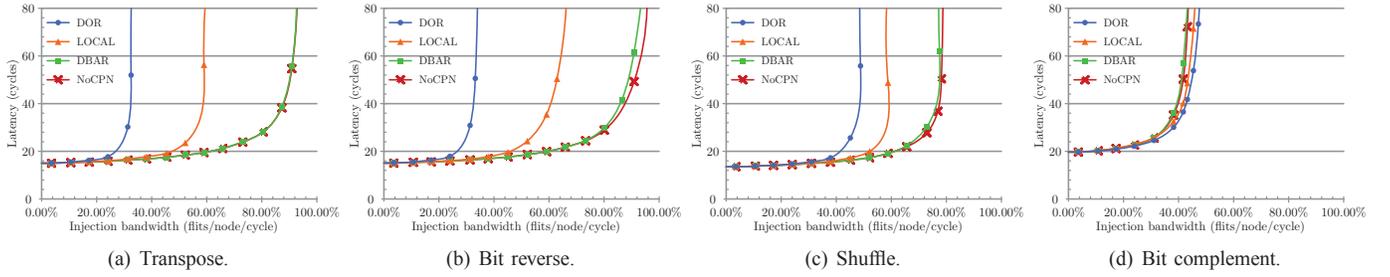,width = \textwidth}
\end{center}
\caption{Routing algorithm performance for 4 x 4 mesh network.}
\label{fig:4x4}
\end{figure*}

\begin{figure*}[htbp!]
\begin{center}
\epsfig{file=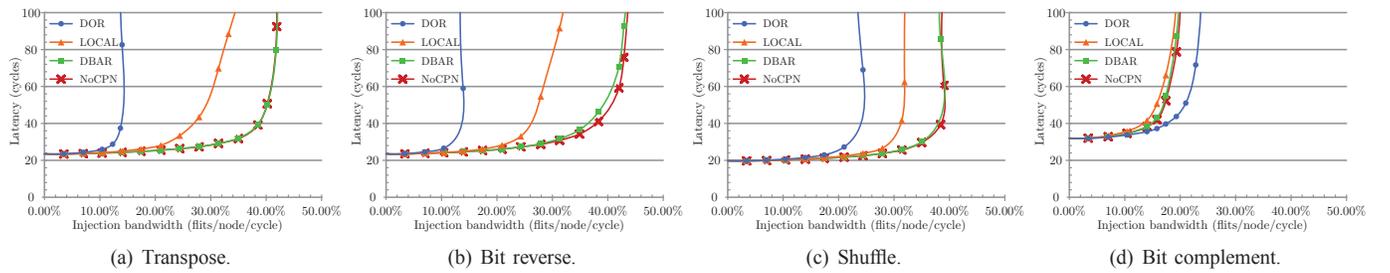,width = \textwidth}
\end{center}
\caption{Routing algorithm performance for 8 x 8 mesh network.}
\label{fig:8x8}
\end{figure*}

As shown in Fig. \ref{fig:4x4} and \ref{fig:8x8}, our algorithm
(NoCPN) have better performance than DBAR on Bit reverse, Shuffle,
Bit complement and have almost the same performance on Transpose.

\bibliographystyle{IEEEbib}
\bibliography{routing}

\end{document}